\documentclass{eas}
\usepackage{graphicx}
\begin{document}

\TitreGlobal{Mass Profiles and Shapes of Cosmological Structures}

\title{The dwarf galaxy DDO 47: testing cusps hiding in triaxial halos}
\author{Gentile, G.}\address{SISSA, Via Beirut 2, 34014 Trieste, Italy}
\author{Burkert, A.}\address{Universit\"ats-Sternwarte M\"unchen, Scheinerstr. 1, 81679 M\"unchen, 
Germany }
\author{Salucci, P.}\address{SISSA, Via Beirut 2, 34014 Trieste, Italy}
\author{Klein, U.}\address{RAIUB, Auf dem H\"ugel 71, 53121 Bonn, Germany}
\author{Walter, F.}\address{Max Planck Institut f\"ur Astronomie, K\"onigstuhl 17, 69117 Heidelberg,
Germany}
\runningtitle{The dwarf galaxy DDO 47: testing cusps hiding in triaxial halos}
\setcounter{page}{23}
\index{Gentile, G.}

%
\begin{abstract} 
We present HI data of the dwarf galaxy DDO 47, aimed at testing
the hypothesis that dark halo triaxiality might induce non-circular motions resulting
in rotation curves best fitted by cored halos, even if the dark matter halo is intrinsically
cuspy. 
We performed a harmonic decomposition of the velocity field in order to search for 
alleged non-circular motions needed to ``hide'' a cusp: in DDO 47
non-circular motions are globally at a level of 2-3 km s$^{-1}$, 
far from being sufficient to 
reconcile the observed rotation curve
with the $\Lambda$CDM predictions.
We conclude that the dark matter halo around 
DDO 47 is truly cored and that a cusp cannot be hidden by non-circular motions.
More details are shown in Gentile et al. (2006, ApJL in press, astro-ph/0506538).
\end{abstract}
\maketitle
%
\section{Introduction}

A fundamental prediction of the cosmological ($\Lambda$) Cold Dark Matter ($\Lambda$CDM) theory is
that virialized dark matter halos should have universal density profiles, such as the
Navarro, Frenk \& White (1996, NFW) halo.
These CDM predictions have been confronted with observations, and
numerous mass models of spiral, dwarf and LSB galaxies have shown that
in a large number of cases the inferred cores of dark matter halos
are much shallower than expected from the NFW-fit
(e.g., de Blok et al. 2001, Gentile et al. 2004).
We have used DDO 47 to explore the possibility that solid-body 
rotation curves (usually interpreted as a signature of constant 
density cores) would arise naturally also in NFW halos as a 
result of a triaxial dark matter
mass distribution that hides the steep central density cusp
through deviations from axial symmetry in the disk.

\section{Results}

DDO 47 is one of the most clear cases with kinematical properties that
are inconsistent with NFW (Fig. 1, left panel).

\begin{figure}[t]
   \centering
   \includegraphics[width=12.0cm]{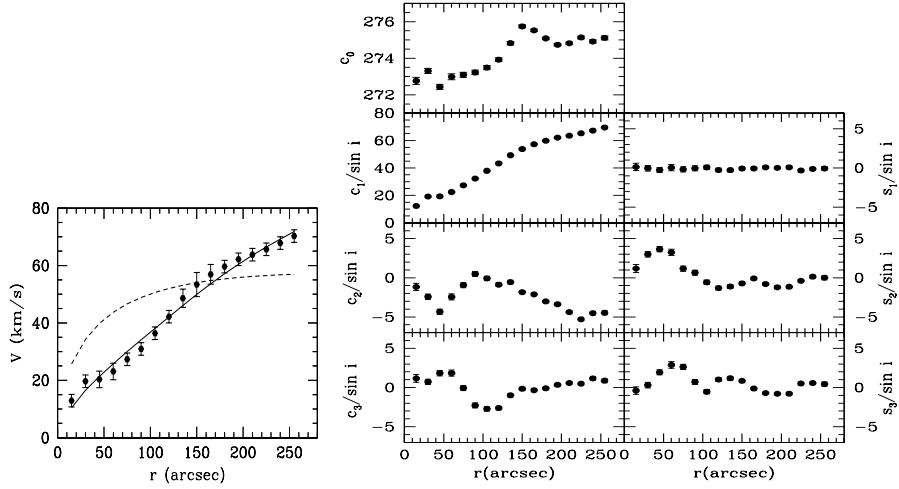}
   \caption{
Left: Mass models for DDO 47: the solid line is the fit with the Burkert halo, and 
the dotted line is the NFW fit. 
Right: harmonic decomposition of the velocity field.}
    \end{figure}

In order to better investigate the kinematic properties of DDO 47, we have fitted 
the velocity field in terms of harmonic coefficients (e.g. Schoenmakers, Franx and de Zeeuw 1997),
with terms up to third order (Fig. 1, right panel). A global elongation of the
potential would cause the $s_3$ term to be approximately constant and non-zero, while 
wiggles could be explained by the presence of spiral structure. In DDO 47 the latter
case applies, we find no evidence for a global elongation of the potential and 
the amplitude of the oscillations is much smaller (a factor 5-10) than the discrepancy
with the $\Lambda$CDM predictions. 
We can conclude that the non-circular motions in DDO 47 are at a level
of at most 3 km s$^{-1}$ and that they are not associated with a global elongation
of the potential, therefore the dark matter halo around 
the dwarf galaxy DDO 47 is truly cored and that a cusp cannot be hidden by non-circular motions.

\end{document}